\documentstyle[11pt,paspconf,epsf]{article}

\newcommand{\bell}{\mbox{\boldmath$\ell$}}
\newcommand{\bT}{\mbox{\boldmath$T$}}

\begin{document}

\title{Nonlinear fluid dynamics of warped discs}
\author{G. I. Ogilvie}
\affil{Institute of Astronomy, University of Cambridge, Madingley Road,
Cambridge CB3 0HA, United Kingdom}

\begin{abstract}
A new description of the dynamics of warped accretion discs is
presented.  A theory of fully nonlinear, slowly varying bending waves
is developed, involving a proper treatment of viscous fluid dynamics
but neglecting self-gravitation.  The special case of resonant wave
propagation found in inviscid Keplerian discs is not considered.  The
resulting description goes beyond existing linear and nonlinear
theories, and is suitable for numerical implementation.
\end{abstract}


\section{Introduction}

Non-planar accretion discs are believed to exist in many astrophysical
situations.  In some cases, such as the nuclear disc of the galaxy NGC
4258 or the circumstellar disc of $\beta$ Pictoris, a warped profile
is observed more or less directly.  In others, including the X-ray
binaries Her X-1 and SS 433, tilting of the disc is proposed in order
to explain certain observational characteristics.  A more complete
discussion is given by Pringle (this volume).

In accretion discs, unlike the case of galactic warps, the effects of
self-gravitation may often be neglected.  The first attempts to
describe the time-dependence of a warped disc within the framework of
viscous fluid dynamics (Petterson 1978; Hatchett, Begelman and Sarazin
1981) indicated that the warp would simply diffuse away on a viscous
time scale.  However, Papaloizou and Pringle (1983) showed that these
theories were oversimplified and failed to conserve angular momentum
as a result of serious internal inconsistencies.

Special complications arise because of the existence of a resonance in
thin discs that are both Keplerian (or very nearly so) and inviscid
(or very nearly so).  A distinction must therefore be made between the
{\it generic (non-resonant) case\/} and the {\it resonant case\/}
which occurs when the two conditions
\begin{equation}
\left|{{\Omega^2-\kappa^2}\over{\Omega^2}}\right|\la
H/r\qquad\hbox{and}\qquad\alpha\la H/r\label{res}
\end{equation}
are both satisfied.  Here $\Omega$ is the angular velocity, $\kappa$
is the epicyclic frequency, $H/r$ is the ratio of the semi-thickness
of the disc to the radius, and $\alpha$ is the dimensionless viscosity
parameter (Shakura and Sunyaev 1973).  Whether the resonant case
occurs in realistic astrophysical situations is a matter of some
uncertainty.  Current estimates suggest that the resonant case is not
relevant to discs in X-ray binaries or active galactic nuclei, but may
apply in protostellar discs.  However, the resonance is delicate and
might be destroyed by the effects of self-gravitation, magnetic
fields, or turbulence.

The non-resonant case was examined by Papaloizou and Pringle (1983),
who considered a Keplerian disc with a significant viscosity.  They
treated the warp as a slowly varying disturbance with azimuthal wave
number $m=1$, using linear Eulerian perturbation theory.  The
resulting equation for the warp is a complex linear diffusion
equation, which has been used in applications (e.g.~Kumar and Pringle
1985).

Other cases were treated by Papaloizou and Lin (1994, 1995), who
considered an inviscid, but not necessarily Keplerian, disc.  Again,
linear Eulerian perturbation theory was used, and the authors assumed
the warp to be a slowly varying normal mode of the disc.  In the
non-Keplerian (non-resonant) case the warp obeys a dispersive linear
wave equation, while in the Keplerian (resonant) case the warp obeys a
non-dispersive linear wave equation.  By considering the effect of a
small viscosity on the inviscid modes, Papaloizou and Lin connected
this theory with that of Papaloizou and Pringle (1983), showing how
the transition occurs between wave-like and diffusive behaviour in
Keplerian discs when $\alpha\approx H/r$.  This theory has also been
used in applications (e.g.~Papaloizou and Terquem 1995).

A difficulty arises with these theories because the Eulerian method is
formally valid only when the tilt angle $\beta(r,t)$ of the warp
satisfies $|\beta|\ll H/r$, a condition violated -- almost by
definition -- by any interesting, observable warp.  Of course, a
linear theory may be derived by any convenient method, and violation
of the above condition does not necessarily imply that the linearized
dynamics ceases to apply, only that the Eulerian method cannot be used
to justify it.  Indeed, it is clear the Eulerian method is basically
not well suited to the description of a warp or even a rigid tilt, and
that some kind of Lagrangian method ought to be used to describe warps
of significant amplitude and to discuss nonlinear effects.  The
Eulerian method offers little information about nonlinear effects,
although it does predict that, in the resonant case, an amplitude
$|\partial\beta/\partial\ln r|\approx H/r$ would result in horizontal
shearing motions in the disc comparable to the sound speed, which are
expected to be unstable (Kumar and Coleman 1993; Gammie, Goodman and
Ogilvie, in preparation).

Meanwhile, Pringle (1992) developed a different approach in which the
forms of the equations governing a warped viscous disc are derived
simply by requiring mass and angular momentum to be conserved, but
without reference to the detailed internal fluid dynamics of the disc.
In this scheme, neighbouring rings in the disc exchange angular
momentum by means of viscous torques which are of two kinds.  One kind
of torque (associated with a kinematic viscosity coefficient $\nu_1$)
acts on the differential rotation in the plane of the disc, and leads
to accretion; the other kind (associated with $\nu_2$) acts to flatten
the disc.  The equations derived (originally by Papaloizou and Pringle
1983) are
\begin{equation}
{{\partial\Sigma}\over{\partial
t}}+{{1}\over{r}}{{\partial}\over{\partial r}}(r\Sigma\bar v_r)=0
\end{equation}
for the surface density $\Sigma(r,t)$, and
\begin{equation}
{{\partial}\over{\partial t}}(\Sigma
r^2\Omega\bell)+{{1}\over{r}}{{\partial}\over{\partial r}}(\Sigma\bar
v_rr^3\Omega\bell)={{1}\over{r}}{{\partial}\over{\partial
r}}\left(\nu_1\Sigma r^3{{{\rm d}\Omega}\over{{\rm
d}r}}\bell\right)+{{1}\over{r}}{{\partial}\over{\partial
r}}\left({\textstyle{{1}\over{2}}}\nu_2\Sigma
r^3\Omega{{\partial\bell}\over{\partial r}}\right)\label{jep}
\end{equation}
for the angular momentum, in the absence of external torques.  Here
$\bar v_r(r,t)$ is the mean radial velocity and $\bell(r,t)$ is the
tilt vector, which is a unit vector parallel to the orbital angular
momentum of the ring.  These equations constitute a straightforward
generalization of the equations used for a flat disc (e.g.~Pringle
1981).

This approach is ostensibly valid for warps of large amplitude and
would therefore represent an advance on the linear theory.  It has
been used in several applications, in particular to identify the
radiation-driven instability (Pringle 1996) and to explore its linear
theory (e.g.~Maloney, Begelman and Pringle 1996) and nonlinear
evolution (e.g.~Pringle 1997).  However, because the equations of
Pringle (1992) are derived somewhat heuristically, without reference
to the detailed internal fluid dynamics of the disc, some doubts
remain over the validity of this method.  For example, has any
internal degree of freedom of the rings been neglected?  Is the
interaction between neighbouring rings purely of the assumed form of
viscous torques?  Are there any nonlinear fluid dynamical effects that
might limit the amplitude of the warp?  How are the viscosity
coefficients $\nu_1$ and $\nu_2$ related?  It is therefore important
to investigate whether the equations of Pringle (1992) can be derived
{\it ab initio\/} from the three-dimensional fluid dynamical
equations, and to understand how they connect to the previously
established linear theory.

An entirely different approach to warps of large amplitude, based on
three-dimensional numerical simulations, is described by Nelson (this
volume).

\section{Outline of the method}

The basis of the approach is to develop a hydrodynamic theory of
unforced bending waves of large amplitude in a thin disc.  This is
most naturally done in the case of a spherically symmetric external
potential, in which a flat disc has no preferred plane of orientation.
Continuous with the zero-frequency rigid-tilt mode, there exist
bending waves -- with azimuthal wave number $m=1$ in linear theory --
that are slowly varying in time and space.  In contrast, most of the
modes in a thin disc vary on a time-scale comparable to $\Omega^{-1}$
and on a length-scale comparable to $H$ (e.g.~Lubow and Pringle 1993).
Therefore these bending waves are special and deserve a special
analysis.

A full account of the analysis has been given elsewhere (Ogilvie
1998).  It will suffice here to explain the various steps involved
with an emphasis on physical interpretation.

\subsection{Warped coordinates}

The limitations of an Eulerian approach to the fluid dynamics of a
warped disc have already been noted.  However, a fully Lagrangian
approach is not suitable for dealing with a differentially rotating
flow.  Therefore a `semi-Lagrangian' method is used, in which the
equations are derived in Eulerian form but referred to a coordinate
system that follows the principal warping motion of the disc.  Warped
spherical polar coordinates $(r,\theta,\phi)$ are defined as follows
(Figure~1).  $r$ is the spherical radial coordinate.  On each sphere
$r={\rm constant}$ the usual angular coordinates $(\theta,\phi)$ are
defined, but with respect to an axis that is tilted to point in the
direction of the unit vector $\bell(r,t)$.  The intention is that the
disc matter on each sphere will lie close to $\theta=\pi/2$.
\begin{figure}
\centerline{\epsfbox{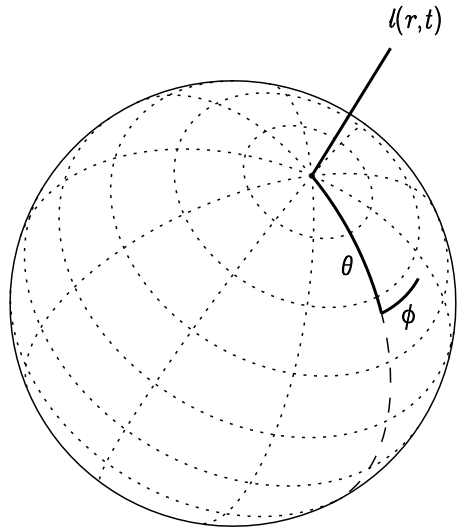}}
\caption{Warped spherical polar coordinates $(r,\theta,\phi)$.}
\end{figure}
This coordinate system is similar to that introduced by Petterson
(1978) but has certain advantages.  It is valid for any amplitude of
warp, whereas Petterson's system, which is based on tilted cylinders,
breaks down for large amplitudes or for insufficiently thin discs.
The present system also has the same Jacobian determinant as ordinary
spherical polar coordinates, which simplifies the equations to some
extent.

The equations of fluid dynamics can be readily derived in these
coordinates using methods of elementary vector calculus.  It is
assumed that the fluid obeys the compressible Navier-Stokes equation
with isotropic viscosity, but viscous heating is neglected.  The
principal features of the equations are (i) the appearance of a number
of fictitious forces and (ii) the introduction of a `modified' radial
derivative operator.  The effect of the latter feature is that in a
warped disc the vertical pressure gradient gives rise to a radial
force which drives horizontal motions (Figure~2).
\begin{figure}
\centerline{\epsfbox{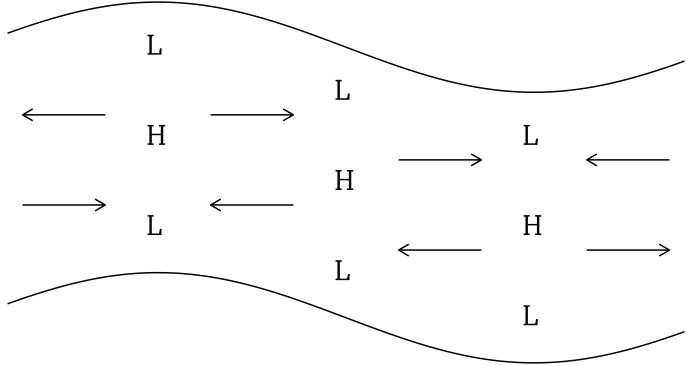}}
\caption{The mechanism by which the stratification of pressure (and
viscous stress) drives horizontal motions in a warped disc.}
\end{figure}

\subsection{Thin-disc asymptotics}

The assumption that the disc is thin allows progress to be made by
introducing scaled variables and asymptotic expansions.  Although
informal thin-disc approximations are common in accretion-disc theory,
the rigorous procedures and wide applicability of asymptotic methods
are not always appreciated.  In the present problem, the basic small
parameter $\epsilon$ is a characteristic value of $H/r$.  The problem
is simplified by (i) separating the `fast' orbital time-scale from the
`slow' time-scale of viscous and dynamical evolution of the warp, (ii)
separating the `fast' orbital velocity, the `intermediate' oscillatory
velocities induced by the warp and the `slow' accretion velocity, and
(iii) allowing thin-disc geometrical simplifications.  For example,
the relative velocity components have the expansions
\begin{eqnarray}
v_r(r,\theta,\phi,t)&=&\epsilon
v_{r1}(r,\phi,\zeta,T)+\epsilon^2v_{r2}(r,\phi,\zeta,T)+O(\epsilon^3),\\
v_\theta(r,\theta,\phi,t)&=&\epsilon
v_{\theta1}(r,\phi,\zeta,T)+\epsilon^2v_{\theta2}(r,\phi,\zeta,T)+
O(\epsilon^3),\\
v_\phi(r,\theta,\phi,t)&=&r\Omega(r)\sin\theta+\epsilon
v_{\phi1}(r,\phi,\zeta,T)+\epsilon^2v_{\phi2}(r,\phi,\zeta,T)+
O(\epsilon^3),
\end{eqnarray}
where $\zeta=\epsilon^{-1}(\pi/2-\theta)$ is a scaled dimensionless
vertical coordinate in the disc, and $T=\epsilon^2t$ is the slow time
coordinate.  Note that all variables are non-axisymmetric except for
the orbital angular velocity $\Omega(r)$.

\subsection{Formal structure of the problem}

After substituting the asymptotic expansions into the dynamical
equations, a number of equations are extracted at different orders.
One of these determines the orbital angular velocity $\Omega(r)$ as
that of a free particle in a circular orbit in the given potential.
The remaining equations may be divided into two sets.  Physically,
Set~A determines the `intermediate' velocities, while Set~B determines
the `slow' velocities.  Mathematically, Set~A consists of coupled
nonlinear partial differential equations (PDEs) in two dimensions
$(\phi,\zeta)$, while Set~B consists of coupled linear PDEs whose
coefficients depend on the solution of Set~A.

It turns out that, while Set~A must be solved in full, all the
information required from Set~B can be extracted by integration.
Indeed, Set~B can be manipulated to derive the basic conservation
equations,
\begin{equation}
{{\partial\Sigma}\over{\partial t}}+{{1}\over{r}}{{\partial}\over{\partial
r}}(r\Sigma\bar v_r)=0,\label{mass}
\end{equation}
\begin{displaymath}
{{\partial}\over{\partial t}}(\Sigma
r^2\Omega\bell)+{{1}\over{r}}{{\partial}\over{\partial r}}(\Sigma\bar
v_rr^3\Omega\bell)={{1}\over{r}}{{\partial}\over{\partial
r}}\left(Q_1{\cal I}r^2\Omega^2\bell\right)+
{{1}\over{r}}{{\partial}\over{\partial r}}\left(Q_2{\cal
I}r^3\Omega^2{{\partial\bell}\over{\partial r}}\right)
\end{displaymath}
\begin{equation}
+{{1}\over{r}}{{\partial}\over{\partial r}}\left(Q_3{\cal
I}r^3\Omega^2\bell\times{{\partial\bell}\over{\partial r}}\right).
\label{am}
\end{equation}
for mass and angular momentum.  Note that this angular momentum
equation is similar to, but more general than, equation (\ref{jep}).
Here ${\cal I}(r,t)$ is the azimuthally averaged second vertical
moment of the density, which in cylindrical polar coordinates would be
written
\begin{equation}
{\cal I}={{1}\over{2\pi}}\int_0^{2\pi}\left(\int_{-\infty}^\infty\rho
z^2\,{\rm d}z\right)\,{\rm d}\phi,
\end{equation}
and is an important dynamical quantity in the theory of bending waves.
The quantities $Q_i$ are dimensionless coefficients which can be
calculated from the solution of Set~A, and which depend on the
amplitude of the warp as well as the rotation law and details of the
thermodynamics and viscosity.  The angular momentum equation implies
that each ring in the disc experiences torques of three kinds from its
neighbours: coefficient $Q_1$ represents a torque tending to spin up
(or down) the ring.  This would be the usual viscous torque
proportional to ${\rm d}\Omega/{\rm d}r$ in a flat disc, but in a
warped disc there is an additional contribution, not proportional to
${\rm d}\Omega/{\rm d}r$, due to a correlation between the radial and
azimuthal velocities induced by the warp; this vanishes in an inviscid
disc because the radial and azimuthal velocities are perfectly out of
phase.  Coefficient $Q_2$ represents a torque tending to align the
ring with its neighbours, which acts to flatten the disc; this also
vanishes in an inviscid disc.  Coefficient $Q_3$ represents a torque
tending to make the ring precess if it is misaligned with its
neighbours; this leads to the dispersive wave-like propagation of the
warp.

\subsection{Separation of variables}

The further assumptions that (i) the disc is polytropic (or
isothermal) locally in radius and (ii) the viscosity coefficients are
locally proportional to the pressure allow Set~A to be solved by
separation of variables, since the vertical dependence of the solution
can be determined by inspection.  The problem is then reduced to
solving a set of dimensionless nonlinear ordinary differential
equations (ODEs) in azimuth, subject to periodic boundary conditions,
and extracting the three dimensionless coefficients $Q_1$, $Q_2$ and
$Q_3$ from the solution.  The system is of fourth order and involves
five dimensionless parameters: the dimensionless amplitude of the
warp,
\begin{equation}
|\psi|=\left|{{\partial\bell}\over{\partial\ln r}}\right|,
\end{equation}
the dimensionless squared epicyclic frequency,
\begin{equation}
\tilde\kappa^2={{{\rm d}\ln(r^4\Omega^2)}\over{{\rm d}\ln r}},
\end{equation}
the polytropic coefficient $\Gamma$ and the dimensionless viscosity
coefficients $\alpha$ and $\alpha_{\rm b}$ (for shear and bulk
viscosities, respectively).

\subsection{Evaluation of the coefficients}

Examples of the numerical evaluation of the coefficients are given by
Ogilvie (1998).  Of particular importance is the diffusion coefficient
$Q_2$; this is shown in Figure~3 for the case of an isothermal,
Keplerian disc without bulk viscosity ($\Gamma=1$; $\tilde\kappa^2=1$;
$\alpha_{\rm b}=0$).  It can be seen that the resonant behaviour
associated with the limit $\alpha\to0$ is diminished as the amplitude
of the warp increases.
\begin{figure}
\centerline{\epsfbox{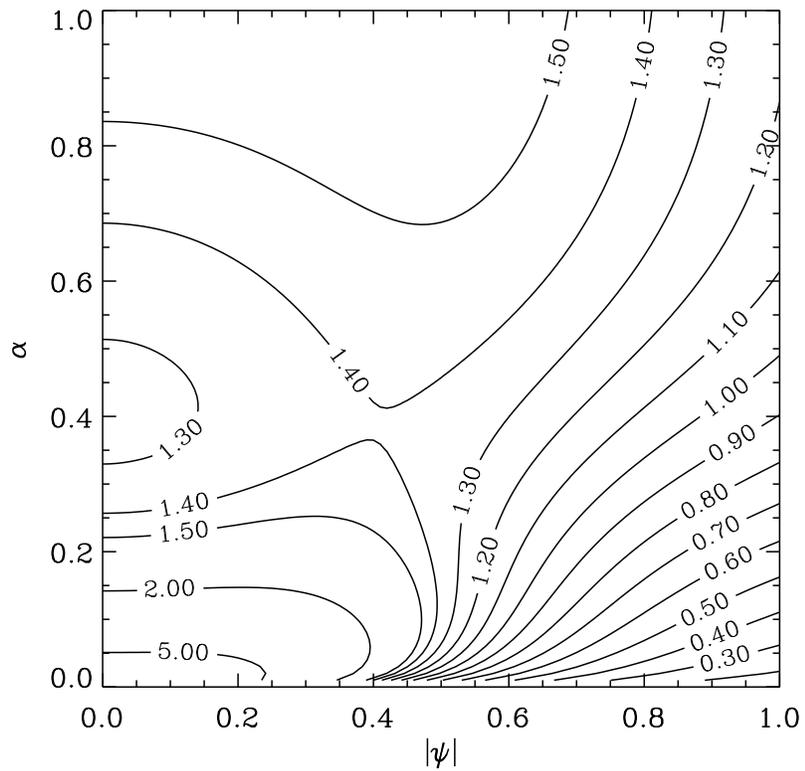}}
\caption{Contour plot of the coefficient $Q_2$ for an isothermal,
Keplerian disc without bulk viscosity.  The horizontal and vertical
coordinates are the dimensionless amplitude of the warp and the
dimensionless viscosity parameter, respectively.}
\end{figure}

\subsection{Nature of the nonlinearity}

A weakly nonlinear theory can be developed analytically (Ogilvie 1998)
which provides truncated Taylor series of the form
$a+b|\psi|^2+O(|\psi|^4)$ for the three coefficients.  The weakly
nonlinear theory therefore contains a cubic nonlinearity which arises
through a three-mode coupling.  The modes involved (Figure~4) are (i)
the `tilt' mode or f mode ($m=1$, of odd symmetry), consisting locally
of a uniform vertical translation of the disc, (ii) an inertial or r
mode ($m=1$, of odd symmetry), consisting of a horizontal epicyclic
motion proportional to $z$, and (iii) an acoustic or p mode ($m=2$, of
even symmetry), consisting of a vertical motion proportional to $z$.
The compressive nature of the third mode explains why the nonlinear
behaviour depends on the thermodynamics and, in principle, on the bulk
viscosity.
\begin{figure}
\centerline{\epsfbox{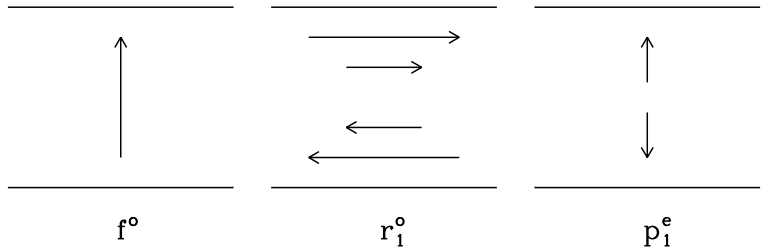}}
\caption{The three modes that couple to produce a cubic nonlinearity.}
\end{figure}

\section{Discussion}

It has been shown that the nonlinear fluid dynamics of a warped
accretion disc is described by conservation equations (\ref{mass}) and
(\ref{am}) for mass and angular momentum, respectively.  The two basic
assumptions are that (i) the disc is thin and (ii) the resonant case
[equation (\ref{res})] does not occur.  Equation (\ref{am}) may be
decomposed into an equation for the component of angular momentum
parallel to $\bell$,
\begin{equation}
\Sigma\bar v_r{{{\rm d}(r^2\Omega)}\over{{\rm
d}r}}={{1}\over{r}}{{\partial}\over{\partial r}}(Q_1{\cal
I}r^2\Omega^2)-Q_2{\cal I}r^2\Omega^2\left|{{\partial\bell}\over{\partial
r}}\right|^2,
\end{equation}
and an equation for the tilt vector,
\begin{displaymath}
\Sigma r^2\Omega\left({{\partial\bell}\over{\partial t}}+\bar
v_r{{\partial\bell}\over{\partial r}}\right)=Q_1{\cal
I}r\Omega^2{{\partial\bell}\over{\partial r}}+
{{1}\over{r}}{{\partial}\over{\partial r}}\left(Q_2{\cal
I}r^3\Omega^2{{\partial\bell}\over{\partial r}}\right)
\end{displaymath}
\begin{equation}
+Q_2{\cal I}r^2\Omega^2\left|{{\partial\bell}\over{\partial
r}}\right|^2\bell+{{1}\over{r}}{{\partial}\over{\partial
r}}\left(Q_3{\cal I}r^3\Omega^2\bell\times{{\partial\bell}\over{\partial
r}}\right).
\end{equation}
The warp therefore satisfies a kind of wave equation of parabolic
type, featuring advection, diffusion and dispersion, each with a
nonlinear dependence on the amplitude of the warp.  This form of
equation arises for the following reason.  According to Set~A, the
velocities induced by the warp are determined instantaneously and
locally in radius.  These velocities are in a `geostrophic' state in
which inertial forces balance the pressure gradients and viscous
forces; time-derivatives of the velocities in the inertial frame do
not appear because of the assumed separation of time-scales.  This
means that the velocities do not constitute additional dynamical
degrees of freedom.  In the resonant case this approximation breaks
down and the time-derivatives must be restored; the resulting
equations for the warp are hyperbolic rather than parabolic.

Since equation (\ref{am}) is the angular momentum equation it is
straightforward to add to the right-hand side any additional torque
$\bT$ due to tidal forcing, Lense-Thirring precession, radiation
forces, self-gravitation, etc.

Under the further assumptions of Section~2.4 concerning the
thermodynamics and viscosity, the coefficients $Q_i$ can be
consistently evaluated.  For small-amplitude warps there is agreement
with linear theory, but for larger amplitudes a numerical solution of
the ODEs is required.

The dispersion coefficient $Q_3$ is generally smaller than the
diffusion coefficient $Q_2$, and, to the extent that it can be
neglected, the angular momentum equation (\ref{jep}) given by Pringle
(1992) is valid.  However, in general, both $\nu_1$ and $\nu_2$ depend
on the amplitude of the warp are neither is equal to the usual
vertically averaged viscosity $\bar\nu$.  The most important result
from linear theory for a Keplerian disc (Papaloizou and Pringle 1983)
was that, although $\nu_1\approx\bar\nu$,
\begin{equation}
{{\nu_2}\over{\nu_1}}\approx{{2(1+7\alpha^2)}\over{\alpha^2(4+\alpha^2)}}
\approx{{1}\over{2\alpha^2}}\qquad\hbox{for $\alpha\ll1$}.
\end{equation}
Therefore, for typical estimates in the range $0.01\la\alpha\la0.1$, a
Keplerian disc is much more resistant to warping than would be
estimated on the basis of the viscosity $\bar\nu$.  The nonlinear
analysis shows that this resonant behaviour is diminished as the
amplitude of the warp increases: the ratio $\nu_2/\nu_1$ decreases
with increasing amplitude of the warp, although it remains
significantly larger than unity.  The consequences of this and other
complex properties of the system ought to be investigated using a
numerical implementation of the governing equations incorporating a
consistent determination of the three coefficients.

\acknowledgments I thank Jim Pringle for many helpful discussions.  I
acknowledge the hospitality of the Isaac Newton Institute during the
programme {\it Dynamics of Astrophysical Discs\/}, where I benefited
from discussions with many participants, and where much of this work
was completed.

\end{document}